# Impressive optoelectronic and thermoelectric properties of two-dimensional $XI_2$ (X=Sn, Si): a first principle study

Atanu Betal, Jayanta Bera and Satyajit Sahu*

*Department of Physics, Indian Institute of Technology Jodhpur, Jodhpur 342037, India*

**Abstract**

Two-dimensional (2D) metal halides have received more attention because of their electronic and optoelectronic properties. Recently, researchers are interested to investigate the thermoelectric properties of metal halide monolayers because of their ultralow lattice conductivity, high Seebeck coefficient and figure of merit. Here, we have investigated thermoelectric and optoelectronic properties of $XI_2$ (X=Sn and Si) monolayers with the help of density functional theory and Boltzmann transport equation. The structural parameters have been optimized with relaxation of atomic positions. Excellent thermoelectric and optical properties have been obtained for both $SnI_2$ and $SiI_2$ monolayers. For $SnI_2$ an indirect bandgap of 2.06 eV was observed and the absorption peak was found at 4.68 eV. For this the highest ZT value of 0.84 for p-type doping at 600K has been calculated. Similarly, for $SiI_2$ a comparatively low indirect bandgap of 1.63 eV was observed, and the absorption peak was obtained at 4.86 eV. The calculated ZT product for $SiI_2$ was 0.87 at 600K. Both the crystals having high absorbance and ZT value suggest that they can be promising candidates for optoelectronic and thermoelectric devices.

**Introduction:**

The increasing demand of energy per capita and increasing population are causing rapid depletion of fossil fuel resources. Renewable energy sources are the only self-sustainable solutions for the advanced world. So, interest has been growing towards research and development of renewable energy sources. In this context thermoelectric (TE) materials have shown promise in which electric power can be generated by a temperature difference across the material owing to Seebeck effect which suggests that if there is some temperature difference between the two ends of the material then it can give rise to a potential difference. Efficiency of thermoelectric materials depends on Seebeck coefficient (S), electrical conductivity (σ), absolute temperature (T) and total conductivity (k). High S, σ value, and low k values are required to get good thermoelectric behavior. After the discovery of graphene, many new 2D materials came into existence [1]–[3]. These 2D materials have shown very interesting properties as compared to bulk materials, such as tunable electronic and optoelectronic properties of materials [4], that led to the use of the materials in energy storage [5], and photovoltaic devices [6][7]. Therefore, people have also shown interest to study and enhance the TE efficiency of 2D transition metal dichalcogenides (TMDC) [8],[9], silicene [10], germanene [11], phosphorene [12]. Because of weak VanderWaals interlayer interaction individual layers of these materials can be exfoliated easily with scotch tape [13]. They can also be grown using chemical vapor deposition (CVD) [14] and thermal evaporation [15] method.

Halide based materials have also attracted researchers because of their excellent semiconducting properties. Among them, group-IV diiodide 2D monolayers are showing great optoelectronic behavior. The stability and electronic property of group-IV iodide 2D materials have been predicted recently [16]. All the monolayer materials have indirect band gap which can be tuned with increase in number of layers. Lead iodide is a good layered material having excellent photoluminescence [17] and electroluminescence [18] property. Good photo absorption and thermoelectric efficiency have also been observed at 300K to 900K in $PbI_2$ [19],[20]. Another same group material $GeI_2$ has also showed excellent thermoelectric and optical properties [21]. These crystals are showing better thermoelectric and optical properties than well-known TMDCs and other 2D materials. For example, $WS_2$ monolayer shows ZT value of 0.7 and 0.4 for n-type and p-type doping respectively at an optimum carrier concentration of $3.5\times10^{20}$ cm$^{-3}$ and at very high temperature of 1500K [22]. $PtSe_2$ monolayer also shows ZT value of 0.65 and 0.25 at 600K [23]. The performance of 2D materials can also be enhanced by applying strain. The improvement of thermoelectric performance of PtSe$_2$ monolayer has also been studied [23]. The strain dependent thermoelectric properties of WS$_2$ has also been studied [24], which shows a better performance at comparatively low temperature of 900K. Large Seebeck value has been seen by antimonene and arsenene but because of large lattice thermal conductivity, ZT value is approaching to 0.8and 0.7 at 700K respectively [25]. Group-III chalcogenides also have been found to be potential candidates as thermoelectric materials. InSe monolayer having hexagonal structure shows high ZT factor of 0.54 (0.48) at 700K for p-type (n-type) doping [28].

Here, for the first time we have studied optoelectronic and low temperature thermoelectric properties of 2D monolayer XI$_2$ (X=Sn and Si) with the help of density functional theory and Boltzmann transport equation. Although various types of 2D materials have been studied but the need of low temperature thermoelectric material was always there. XI$_2$ are group-IV diiodide materials in which people have already investigated thermoelectric properties of PbI$_2$ but the main problem with PbI$_2$ is the toxicity present in lead. In this work we have calculated high thermoelectric figure of merit for SnI$_2$and SiI$_2$at 600K. At room temperature the maximum ZT is 0.66 (0.35), 0.78 (0.51) for p-type (n-type) SnI$_2$ and SiI$_2$ respectively. So, the theoretical investigation of thermoelectric and optical properties has great significance to understand the physical/chemical properties that enhances the thermoelectric efficiency of these materials. These can be used as good thermoelectric materials and used to fabricate devices with excellent efficiency. The absorption peaks occur at ultraviolet region for both materials. So, one of the promising devices that can be fabricated from these materials is a photodetector.

**Methodology:**

The first principle calculations have been carried out using density functional theory (DFT) with Vanderbilt ultrasoft pseudopotential [29] and Perdew-Burke-Ernzerh (PBE) as generalized gradient approximation (GGA)[30] in Quantum Espresso (QE) package[31]. To avoid periodic boundary approximation and interaction between two layers a sufficient vacuum of 20 Å has been kept along Z direction. The calculations have been performed with 15×15×1 k-mesh sampling with ordinary Gaussian spreading. The atoms were relaxed to their equilibrium position until a force convergence threshold of 10$^{-3}$ev/Å has been achieved and hence helped in optimizing the cell parameters. A 45×45×1 k-mesh was considered for optical calculation and on-self-consistent calculation. All the calculations were carried out with 45 Ry cut-off energy and convergence threshold energy for self-consistency was kept at less than $10^{-9}$ Ry. Phonon band structure was calculated with Density Functional Perturbation Theory (DFPT) as implemented in QE with 6×6×1 q-mesh grid. The optical properties were calculated by using the Time Dependent Density Functional Perturbation Theory (TD-DFPT) with SIESTA package [32]. Imaginary and real parts of dielectric function were optimized from momentum space formulation and Kramers-Kronig transformation [33] respectively. From dielectric function other optical properties like absorption coefficient ($\alpha$), refractive index (η), extinction coefficient (K) can be obtained from the following equations.

$$\eta = \left[\frac{\{(\varepsilon_1^2+\varepsilon_2^2)^{1/2}+\varepsilon_1\}}{2}\right]^{1/2} \quad (1)$$

$$K = \left[\frac{\{(\varepsilon_1^2+\varepsilon_2^2)^{1/2}-\varepsilon_1\}}{2}\right]^{1/2} \quad (2)$$

$$\alpha = \frac{2K\omega}{C} \quad (3)$$

Where, $\varepsilon_1, \varepsilon_2, \omega, C$ are the real and imaginary functions of dielectric, frequency of incident light, speed of light. Thermoelectric parameters were calculated from Boltzmann transport equation and constant scattering time approximation as implemented in BoltzTraP code [34]. The transport properties can be obtained from the following equations

$$\sigma_{l,m} = \frac{1}{\Omega}\int \sigma_{l,m}(\varepsilon)\left[-\frac{\partial f_\mu(T,\varepsilon)}{\partial \varepsilon}\right]d\varepsilon \quad (4)$$

$$k_{l,m}(T,\mu) = \frac{1}{e^2 T\Omega}\int \sigma_{l,m}(\varepsilon)(\varepsilon-\mu)^2\left[-\frac{\partial f_\mu(T,\varepsilon)}{\partial \varepsilon}\right]d\varepsilon \quad (5)$$

$$S_{l,m}(T,\mu) = \frac{(\sigma^{-1})_{n,l}}{eT\Omega}\int \sigma_{n,m}(\varepsilon)(\varepsilon-\mu)\left[-\frac{\partial f_\mu(T,\varepsilon)}{\partial \varepsilon}\right]d\varepsilon \quad (6)$$

Where, $\sigma_{l,m}, k_{l,m}, S_{l,m}$ e, T, $\Omega$, μ are the electrical conductivity, electronic thermal conductivity, Seebeck coefficient, electron charge, absolute temperature, volume of the unit cell, chemical potential respectively. In terms of group velocity conductivity tensor can be obtained from the following equation.

$$\sigma_{l,m}(i,\boldsymbol{k}) = e^2 \tau_{i,\boldsymbol{k}} v_l(i,\boldsymbol{k}) v_m(i,\boldsymbol{k}) \quad (7)$$

Where, $v_l$ is the group velocity, given by

$$v_l(i, \boldsymbol{k}) = \frac{1}{\hbar} \frac{\partial \varepsilon_{i,\boldsymbol{k}}}{\partial \boldsymbol{k}_l} \tag{8}$$

Where, $\boldsymbol{k}_l$ is the $l^{th}$ component of wave vector $\boldsymbol{k}$ and $\varepsilon_{i,\boldsymbol{k}}$ is the energy of the $i^{th}$ band.

Lattice thermal conductivity ($k_{ph}$) and phonon lifetime of materials were calculated using supercell approach with phono3py package [35] interfaced with QE. Supercell of 2×2×1 with k-mesh of 6×6×1 was constructed and self-consistent calculations of supercells with finite displacement of 0.06 Å have been included to calculate $k_{ph}$ and lifetime.

**Results and Discussions:**

**Structural properties and mechanical stability:**

$XI_2$ (X=Sn and Si) monolayer has a hexagonal honeycomb like structure with space group 164 (P-3m1). The monolayer has three atomic layers with X as the middle layer and the I atoms constitute the upper and lower atomic layers respectively. So, a unit cell is consisting of one X atom and two I atoms. The side and the top views of the unit cell of $XI_2$ are shown in fig. 1(a) and (b) respectively. Each X atom is surrounded by six I atoms and the I atoms themselves form closed heaxagonal structure. The lattice constants were optimized by relaxing the unit cell. For $SnI_2$ monolayer, the optimized lattice constants a=b=4.52Å and height 3.73Å were in good agreement with previous experimental results[15]. In each unit cell two different angles 89.27° and 89.0° for ∠I-Sn-I and ∠Sn-I-Sn respectively were obtained. Similarly, for $SiI_2$ the corresponding optimized lattice constants, angles ∠I-Si-I, and ∠Si-I-Si after relaxation are a=b= 4.19Å, 90.86°, 89.93° respectively. The distances and angles between atoms are shown in table-1.

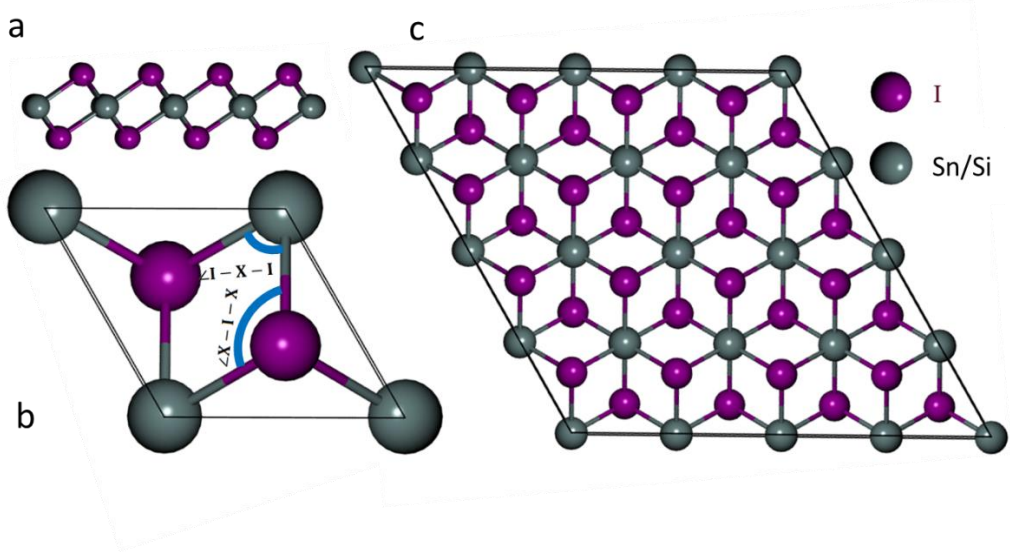

**Figure1.** (a) Side view of XI$_2$ monolayer. (b) The top view of the unit cell which also shows the angles ∠I-X-I and ∠X-I-X. (c) The top view of XI$_2$super cell, which clearly shows the honeycomb structure.

| Crystal | a=b (Å) | d$_{X-I}$ (Å) | d$_{I-I}$ (Å) | Height (Å) | $\theta_{I-X-I}$ | $\theta_{X-I-X}$ | Coheshive Energy (eV/atom) |
|---|---|---|---|---|---|---|---|
| SnI$_2$ | 4.52 | 3.2249 | 4.5302 | 3.73 | 89.27° | 89.01° | 14.286 |
| SiI$_2$ | 4.19 | 2.9837 | 4.1921 | 3.43 | 90.84° | 89.93° | 2.39 |

**Table 1.** Stuctural information of SnI$_2$, and SiI$_2$ monolayers.

The cohesive energy calculation of the unit cells gives us the information about the stability of the unit cell which can be obtained by using the following equation: $E_{ch} = \{(E_X + 2.E_I) - E_{XI_2}\}/3$, where $E_{XI_2}$, $E_X$, $E_I$ are the total energy of 2D monolayer, isolated X atom, and isolated I atom respectively. Using this we found out the cohesive energy per atom in the 2D monolayer SnI$_2$ and SiI$_2$ are 14.286 eV and 2.39 eV respectively, which gives us the information that these structures are stable.

To check the thermodynamical stability due to the vibration of the atoms in the lattice the phonon dispersion data along the high symmetry path Γ-M-K-Γ were studied and is shown for SnI$_2$ and SiI$_2$ in fig. 2(a) and (b) respectively. The dynamic stability of the structures were confirmed from the absence of the imaginary frequencies. Total nine vibrational modes were obtained due to the three atoms in the unit cell, out of which three are acoustic modes and six are optical modes obtained due to the in-phase and out of-phase vibration of the atoms respectively in the unit cell. The three aoustic modes are in plane transverse (TA), and longitudinal (LA) modes, and out of plane mode (ZA). The highest frequency of the optical mode is 150 cm$^{-1}$ for $SnI_2$ and there is a finite gap between the optical and the acoustic modes. For $SiI_2$, the highest frequency for the optical mode is 260 cm$^{-1}$ but in this case the optical and the acoustic modes are overlapping with each other which indicates the scattering of the optical and acoustic vibrations.

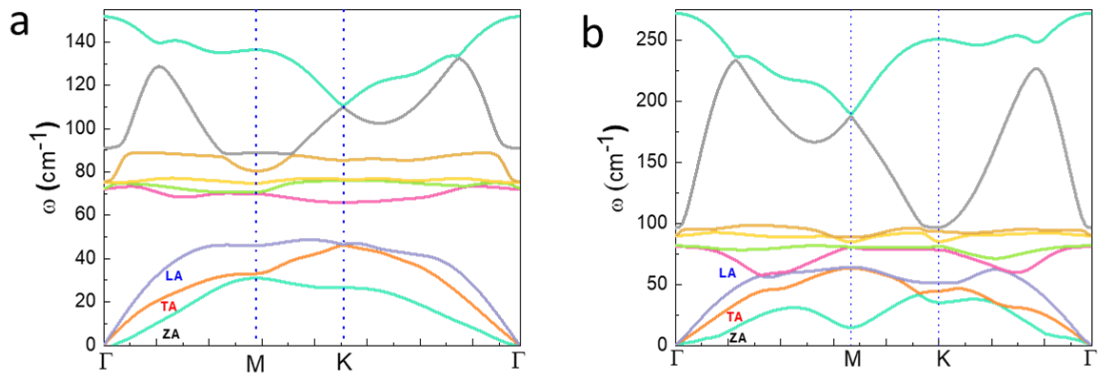

**Figure 2.** Phonon dispersion curve of (a) $SnI_2$ monolayer and (b) $SiI_2$ monolayer. The acoustic in-pane transverse, longitudinal modes and the out of plane mode are shown as TA, LA, and ZA respectively.

**Electronic properties:**

The electronic band structure of $XI_2$ monolayer has been studied within the energy range of -5 eV to 5 eV in the first Brillouin zone along the K-Γ-M-K path. Fig. 3(a) and (b) show the band structure of $SnI_2$ and $SiI_2$ and the corresponding bandgaps are 2.06 eV and 1.63 eV respectively which match well with the previous data [16]. Both the structures show indirect bandgap. For $SnI_2$ and $SiI_2$ the valence band maxima (VBM) lies somewhere between the symmetry point K and Γ, whereas the conduction band minima (CBM) lies at Γ point. The valence band has two almost degenerate energy states lying along the K-Γ and Γ-M paths. This is because of the presence of two I atoms.

The projected density of states (PDOS) of $SnI_2$ in fig. 3(c) indicates that the contribution toward VBM is mainly due to the $p_x$, $p_y$ and $p_z$ orbitals of I atoms and s orbital of Sn atom. Similarly, the contribution toward CBM comes mainly from $p_x$, $p_y$ and $p_z$ orbitals of Sn atom whereas the contribution of p-orbitals of I atoms and s-orbital of Sn atom is negligible. Similarly, for $SiI_2$ as shown in fig. 3(d) the contribution toward VBM comes mainly from $p_x$, $p_y$ and $p_z$ orbital of I atom and s orbital of Si atom and the contribution toward CBM comes mainly from $p_x$, $p_z$ orbitals of Si atom.

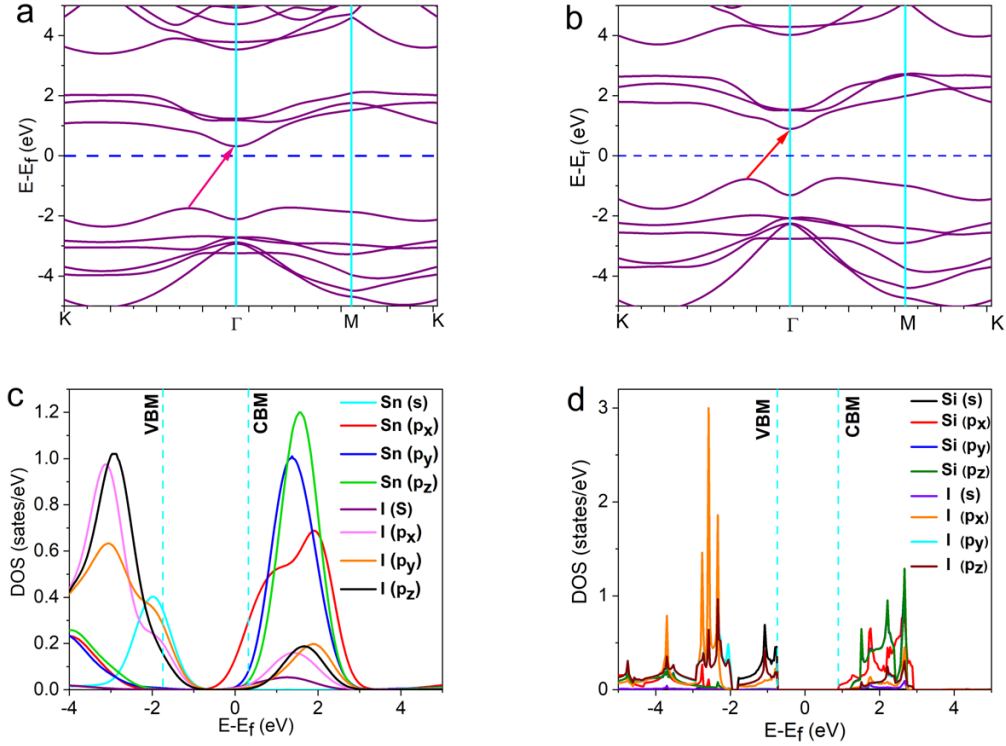

**Figure 3**. Band structure of (a) $SnI_2$ monolayer and (b) $SiI_2$ monolayer. Projected density of state (PDOS) of (c) $SnI_2$ & (d) $SiI_2$ monolayer.

**Carrier mobility and relaxation time:**

Mobility of charge carriers is calculated by using Bardeen and Shockley's deformation potential theory, in which shift in band edge with applied strain can be referred to as deformation potential of the crystal [36]. The expression of mobility for 2D system can be derived from deformation potential and effective mass theorem given as

$$\mu_{2D} = \frac{e\hbar^3 C_{2D}}{K_B T m_e^* m_d (E_l^2)} \qquad (9)$$

Here, T is the temperature in Kelvin, $m^* = \frac{d^2E}{dk^2}$ is the effective mass of charge carrier which can be found out from band edge of band structure and average effective mass $m_d$ can be determined as $m_d = \sqrt{m_x^* m_y^*} \cdot \frac{d^2E}{dk^2}$ can be determined by fitting the parabola of band edge of VBM for hole and CBM for electron. $E_I = \frac{\partial V}{\partial \varepsilon}$ is the deformation potential constant of the material, which is calculated from the slope of energy of band edges (V) with applied strain ($\varepsilon$). $C_{2D} = (\partial^2 E / \partial \varepsilon^2)/A_0$ is the elastic constant of 2D monolayer, $\varepsilon$ is percentage of strain applied to the system, $A_0$ is the area of unstrained system. Relaxation time of the charge carrier can be calculated by using the well-known formula $\tau = \frac{\mu m^*}{e}$. All calculated parameters for both the crystals are listed in table 2.

| material | $E_{I,electron}$ (eV) | $E_{I,hole}$ (eV) | $C_{2D}$ (J/m²) | $m_e^*$ (Kg) | $m_h^*$ (Kg) | $\tau_e$ (s) | $\tau_h$ (s) | $\mu_e$ (cm²V⁻¹s⁻¹) | $\mu_h$ (cm²V⁻¹s⁻¹) |
|---|---|---|---|---|---|---|---|---|---|
| SnI$_2$ | 13.684 | 14.674 | 87.97 | 0.271 m₀ | 0.337 m₀ | 2.08×10⁻¹⁴ | 1.45×10⁻¹⁴ | 135.0 | 76.05 |
| SiI$_2$ | 16.438 | 17.048 | 96.06 | 0.254 m₀ | 0.474 m₀ | 1.66×10⁻¹⁴ | 0.83×10⁻¹⁴ | 115.75 | 31.05 |

**Table 2.** Calculated deformation potential constant, elastic constant, effective mass, relaxation time, mobility of SnI$_2$ and SiI$_2$ monolayer.

**Optical properties:**

Optical properties of XI$_2$ (X=Sn and Si) 2D monolayers were calculated along the perpendicular direction of the plane. The imaginary part of dielectric constant($\varepsilon_2$) is calculated using momentum space formulation with proper matrix elements, whereas real part of dielectric constant($\varepsilon_1$) has been obtained from Kramers-Kronig transformation. The dielectric function vs energy of incident photon is shown in fig. 4(a). Imaginary part ($\varepsilon_2$) for SiI$_2$ starts increasing at 1.1 eV and first peak was observed at 3.59 eV with value 4.55, whereas for SnI$_2$ it started increasing at 1.6 eV and the peak was observed at 3.78 eV with maximum value of 4.37. The energy gap between the two can be explained from the band structure. SnI$_2$ is having bandgap of 2.06 eV which is 0.43 eV greater than SiI$_2$ bandgap resulting a blue shift of the bandgap in SnI$_2$. A secondary peak is observed at 8.09 eV and 8.55 eV for SnI$_2$ and SiI$_2$ respectively. The real dielectric constant ($\varepsilon_1$) is finite at zero energy which is 3.13 and 3.85 for SnI$_2$, and SiI$_2$ respectively. The peaks are obtained at 2.16 eV and 2.54 eV for SnI$_2$ and SiI$_2$ respectively. A negative value for $\varepsilon_1$ can also be seen at 4.59 eV and 5.94 eV which becomes constant after 15 eV.

The absorption spectra are shown in fig. 4(b) which is similar to the imaginary dielectric constant in lower energy region. The first absorption peak was seen at 4.68 eV and 4.86 eV with $\alpha$ value of $5.97\times10^5$ cm$^{-1}$ and $6.11\times10^5$ cm$^{-1}$ for SnI$_2$ and SiI$_2$ respectively, which are of same order as that of GeI$_2$ [21]. Some secondary peaks are also observed at higher photon energy with lower absorbance and above 25 eV the absorbance is negligible. Hence both the materials can be used in optoelectronic devices. Since the absorption peak is in the UV region hence the materials can be used in UV photodetector. The refractive index with respect to energy is shown in fig. 4(c) which is 1.77 at zero energy and at 2.69 eV it gives the highest value of 2.21 for SiI$_2$, whereas for SnI$_2$ the value is 1.96 at zero energy and 2.32 at 2.34 eV. A secondary peak arises at 7.56 eV and 7.33 eV for SiI$_2$ and SnI$_2$ respectively. Both the materials have almost constant refractive index of 0.8 after 15 eV.

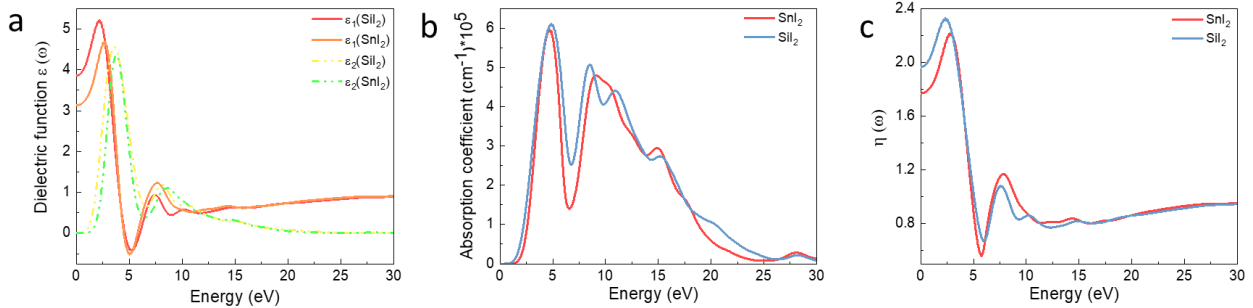

**Figure 4.** (a) The real part of dielectric function for monolayer SnI$_2$ (solid red line) and SiI$_2$ (solid orange line) and imaginary part of dielectric function of SnI$_2$ (dotted green line) and SiI$_2$ (dotted yellow line). (b) Absorption coefficient plotted with photon energy of SnI$_2$ (red line) and SiI$_2$ (blue line). (c) Refractive index of SnI$_2$ (red line) and SiI$_2$ (blue line).

**Thermoelectric performance:**

One of the important thermoelectric properties to assess the thermoelectric performance of the material is Seebeck coefficient (S). The variation of Seebeck coefficient with chemical potential for SnI$_2$ is shown in fig. 5(a). The value of S for p-type and n-type carriers in SnI$_2$ at 300K is 2930μV/K and 2629 μV/K respectively. With the increase in temperature the value of S decreases because S is inversely proportional to the absolute temperature T. The variation of relaxation time scaled electrical conductivity with chemical potential is shown in fig. 5(b) which hardly changes with the increase in temperature. The electrical conductivity is high for p-type carriers which is $0.91\times10^{19}$ S/m in comparison to n-type charge carriers which is $0.17\times10^{19}$ S/m at the band edges in SnI$_2$ monolayer. The relaxation time scaled power factor (PF) which is shown in fig. 5(c) is increasing with increase in temperature. The highest value of PF for p-type and n-type carriers at 600K is $21.17 \times 10^{10}$ W/m-K$^2$-s and $7.63 \times 10^{10}$ W/m-K$^2$-s respectively. The higher value of S, $\sigma/\tau$, $S^2\sigma/\tau$ for p-type carriers in comparison to n-type carriers suggest that SnI$_2$ will be effective as a TE material when doped with p-type material. The S of monolayer SiI$_2$ at different temperatures are shown in fig. 5(d). The values of S at 300K for p-type and n-type carrier are 2596 μV/K and 2556μV/K respectively and which are higher than that of SnI$_2$. This can be explained from the perspective of bandgap which is higher for SnI$_2$ which means there are less number of thermally excited charge carriers near the band edges and Seebeck coefficient is directly proportional to charge. PF for SiI$_2$ is higher for p-type charge carriers with its maximum value of 23.41 W/m-K$^2$-s. It is observed that the PF is not increasing significantly after 600K for both SnI$_2$ and SiI$_2$. So, one can use these materials as low temperature thermoelectric materials. The values of S, $\sigma/\tau$, $S^2\sigma/\tau$, and ZT product at different temperatures for SnI$_2$ and SiI$_2$ are shown in table 3.

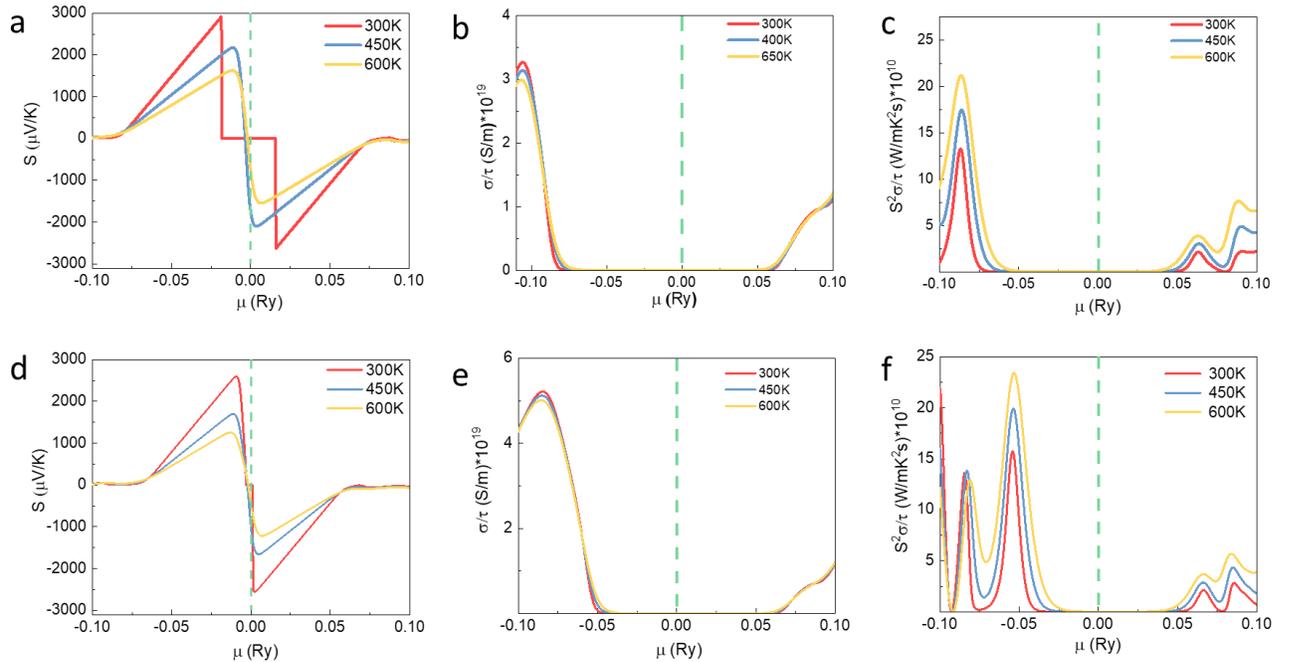

**Figure 5.** Thermoelectric parameters for SnI$_2$ are shown in figure (a) to (c) where (a) shows variation of Seebeck coefficient (b) shows variation of $\sigma/\tau$ and (c) variation of $S^2\sigma/\tau$ with chemical potential(μ) at different temperatures. Similarly, for SiI$_2$ the variation of (d) Seebeck coefficient (e) $\sigma/\tau$ and (f) $S^2\sigma/\tau$ with chemical potential (μ) at different temperatures.

| Material | | S in (µV/K) | | σ/τ*10¹⁹ (S/m) | | S²σ/τ *10¹⁰ (W/mK²s) | | ZT | |
|---|---|---|---|---|---|---|---|---|---|
| | | P-type | n-type | P-type | n-type | P-type | n-type | P-type | n-type |
| SnI$_2$ | 300K | 2930 | 2629 | 3.26 | 5.86 | 13.26 | 3.89 | 0.66 | 0.35 |
| | 450 K | 2176 | 2098 | 3.13 | 5.75 | 17.44 | 4.89 | 0.78 | 0.55 |
| | 600 K | 1623 | 1545 | 2.98 | 5.61 | 21.17 | 7.63 | 0.84 | 0.65 |
| SiI$_2$ | 300 K | 2596 | 2556 | 5.21 | 7.85 | 15.73 | 4.07 | 0.78 | 0.51 |
| | 450 K | 1700 | 1658 | 5.12 | 7.72 | 19.90 | 6.43 | 0.84 | 0.66 |
| | 600 K | 1255 | 1215 | 5.01 | 7.58 | 23.41 | 9.67 | 0.87 | 0.74 |

**Table 3.** The calculated values of Seebeck coefficient (S), relaxation time scaled electrical conductivity (σ/τ) and power factor (S²σ/τ), thermoelectric figure of merit (ZT) at 300K, 450K, 600K for both p-type and n-type carriers in SnI$_2$ and SiI$_2$ monolayers

**Lattice thermal conductivity:**

Lattice thermal conductivity ($k_{ph}$) of monolayer XI$_2$ has been calculated using linear phonon Boltzmann transport equation (LBTE) and relaxation time approximation as implemented in phono3py package. The variation of lattice thermal conductivity with temperature in SnI$_2$ is shown in fig. 6(a) and the calculated value of $k_{ph}$ is 0.16 W/m-K at 300K. This value is of the same order as that of observed in PbI$_2$ and GeI$_2$[21],[37]. The lattice thermal conductivity as a function of temperature in SiI$_2$ is shown in fig. 6(d) and at 300K the value is 0.05 W/m-K. The variation of thermal conductivity with temperature also follows the same trend as that in SnI$_2$. The lattice thermal conductivity starts decreasing with increase in temperature. This can be understood from Leibfried and Schlomann's model for lattice thermal conductivity which was modified by Slack [38]. The lattice thermal conductivity for non-metallic crystal is given by

$$k_{ph} = A \frac{\bar{m}\Theta_D^3 \delta}{\gamma^2 N^{2/3} T} \qquad (10)$$

Where, A is a constant, $\bar{m}$ is average mass per atom in a crystal, $\Theta_D$ is Debye temperature, $\delta$ is the average volume occupied by one atom, N is number of atoms per unit cell, γ is Gruneisen's constant, the measure of anharomnicity of the crystal[39]. $k_{ph}$ is inversely proportional to temperature which suggests that with increase in temperature the lattice thermal conductivity decreases. The variation of phonon life time and group velocity of SnI$_2$ with frequency is shown in fig. 6(b) and (c)respectively. The maximum lifetime for phonon is 0.7ps which is much smaller than that of MoS$_2$, WS$_2$ and other 2D materials [40], [41]. The maximum phonon group velocity for longitudinal acoustic modes and optical modes is 2.1 km/s and 4.07 km/s respectively. The phonon life time and the phonon group velocity of SiI$_2$ are shown in fig. 6(e) and (f). The maximum phonon life time in SiI$_2$ is 0.3ps, which is smaller than that of SnI$_2$. Similarly, the maximum phonon group velocities for the acoustic and optical modes are 3.12 km/s and 12.42 km/s. The lower thermal conductivity can be explained from the point of view of phonon life time, phonon group velocity, and the phonon dispersion relationship as the decrease in the first two causes a decrease in the thermal conductivity. The thermal conductivity of SiI$_2$ is much smaller than that of SnI$_2$ and this can be easily related with the comparative life time, and group velocity parameters as for SiI$_2$ the values for these two are very small. The phonon dispersion curve also suggests that in case of SiI$_2$, there is an overlapping between the optical and acoustic modes which causes even more scattering of the phonons and hence lesser transport of the phonons leading to decrease in thermal conductivity.

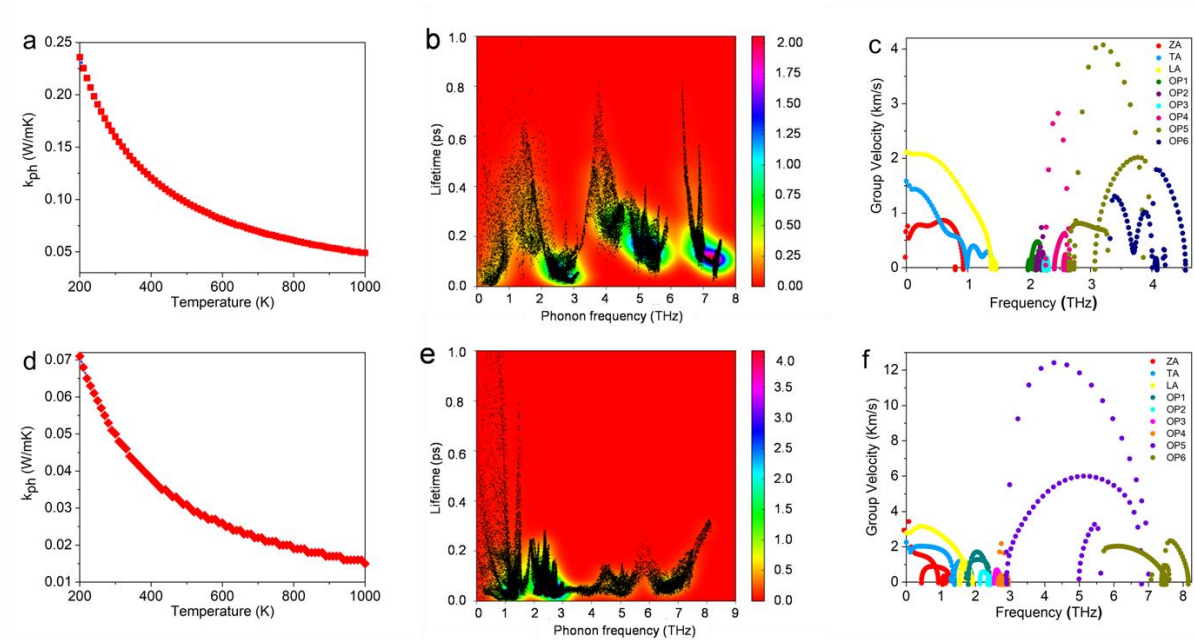

**Figure 6.** For monolayer SnI$_2$ (a) The variation of lattice thermal conductivity with temperature. (b) Variation of phonon life time with frequency. (c) Variation of group velocities of various acoustic and optical modes with frequency in monolayer SnI$_2$. Similarly, for monolayer SiI$_2$ (d) The variation of lattice thermal conductivity with temperature. (e) Variation of phonon life time with frequency. (f) Variation of group velocities of various acoustic and optical modes with frequency.

**Thermoelectric figure of merit (ZT):**

The efficiency of a thermoelectric material called thermoelectric figure of merit (ZT) is calculated by using the parameters we have discussed so far which is related to them as by the following equation:

$$ZT = \frac{S^2 \sigma T}{k} \qquad (11)$$

Where S is Seebeck coefficient, σ is the electrical conductivity, T is absolute temperature and k is total thermal conductivity, the sum of both electronic thermal conductivity and lattice thermal conductivity ($k=k_{el}+k_{ph}$). The variation of ZT with chemical potential of $SnI_2$ and $SiI_2$ at different temperatures are shown in fig. 7(a) and (b) respectively. Two distinct peaks are observed for each temperature. The peaks in the negative and positive chemical potentials (μ<0 and μ>0) are corresponding to the ZT product for p-type and n-type doping respectively. The gap between the two peaks is identical to the energy gap for the corresponding material and in this gap the ZT value is zero. The ZT maxima for the p-type doped $SnI_2$ are 0.66, 0.78, and 0.84 at 300K, 450K, 600K respectively and for n-type doped $SnI_2$ it is 0.35, 0.55, and 0.65 at the same corresponding temperatures respectively. These ZT values are much greater than many commonly studied 2D materials [22],[25]. Similarly, for $SiI_2$ the ZT value maxima for the p-type doped material are 0.78, 0.84, 0.87 and 0.51, 0.66, and 0.74 for n-type doped materials at 300K, 450K, and 600K respectively. As the temperature increases the ZT value also increases but after 600K the value starts to saturate. High ZT product at room temperature suggests that the materials can be used as room temperature thermoelectric material. It is also seen that both the materials have high ZT product when they are doped with p-type materials which suggests that p-type materials behave as better thermoelectric materials. When we look at the thermoelectric properties of both $SnI_2$ and $SiI_2$ we see that $SiI_2$ is a better thermoelectric material than $SnI_2$.

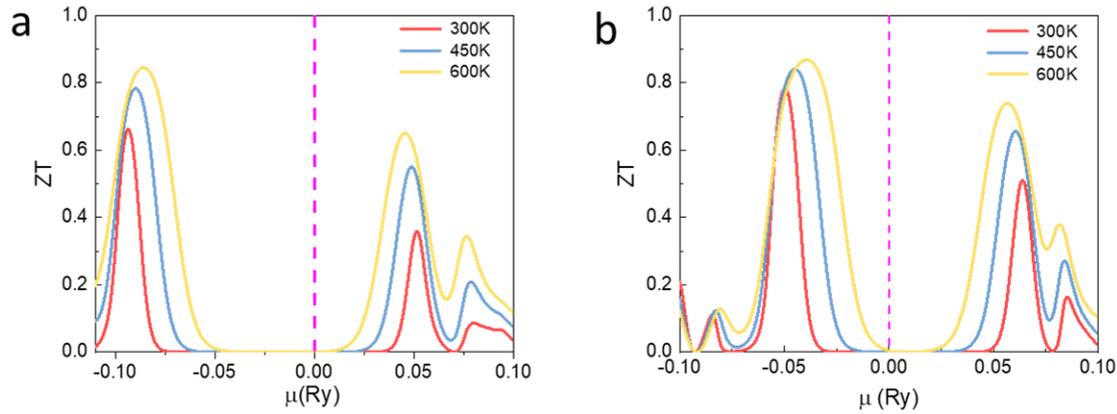

**Figure 7.** Variation of the thermoelectric figure of merit with chemical potential of material for (a) SnI$_2$ monolayer (b) SiI$_2$ monolayer.

**Conclusion:**

We have calculated the electronic, optical, and thermoelectric properties of SnI$_2$ and SiI$_2$ monolayers. Structural parameters were optimized and the dynamical stability was confirmed by studying the cohesive energy and phonon dispersion curve. Both the materials have indirect bandgap with the values of 2.06eV and 1.63eV for SnI$_2$ and SiI$_2$ respectively which match with the reported data. The optical properties were calculated and the absorption peak lies in the UV region which prompts that the materials can be used in UV photodetectors. The thermoelectric properties of the materials are impressive having highest power factor of 21.17 ×10$^{10}$ W/m-K$^2$-s and 23.41×10$^{10}$ W/m-K$^2$-s at 600K for p-type doping for SnI$_2$ and SiI$_2$ monolayers respectively. Both the materials show very low lattice thermal conductivity which leads to high ZT product of the value close to the unity, but out of these two materials SiI$_2$ is a better thermoelectric material because of its lower value of lattice thermal conductivity than that of SnI$_2$. So, both the materials can be used for high performance thermoelectric device fabrication as well as fabrication of photodetectors in the UV region.

**Acknowledgement:**

The authors are thankful to Ministry of Human Resource and Development (MHRD) for providing the funding to carry out the research and to Indian Institute of Technology for providing the infrastructure to conduct the research.